# Using Warrants As a Window to Epistemic Framing

Thomas J. Bing and Edward F. Redish

*Department of Physics, University of Maryland, College Park, MD 20742, USA*

**Abstract.** Mathematics can serve many functions in physics. It can provide a computational system, reflect a physical idea, conveniently encode a rule, and so forth. A physics student thus has many different options for using mathematics in his physics problem solving. We present a short example from the problem solving work of upper level physics students and use it to illustrate the epistemic framing process: "framing" because these students are focusing on a subset of their total math knowledge, "epistemic" because their choice of subset relates to what they see (at that particular time) as the nature of the math knowledge in play. We illustrate how looking for students' warrants, the often unspoken reasons they think their evidence supports their mathematical claims, serves as a window to their epistemic framing. These warrants provide a powerful, concise piece of evidence of these students' epistemic framing.

## INTRODUCTION

Mathematics is central to learning physics. As physics students progress to higher and higher levels of physics classes, the use of mathematics becomes evermore complex. Certainly the mathematics becomes more complex in a conceptual sense—upper level physics uses increasingly technical math. This paper, however, focuses on a different aspect of the complexity of mathematics in physics classes: epistemic complexity. Math in physics can have many different natures: it can be a computational scheme, reflect a physical idea, conveniently encode a rule, and so forth.

This paper will use the idea of epistemic framing to model how physics students juggle these various natures of their mathematics. We will examine the warrants these students use in their arguments as a primary source of evidence for their epistemic framing.

## EPISTEMIC FRAMING

We turn first to a brief overview of two cognitive modeling ideas: epistemic resources and framing. These ideas will be applied to a sample episode of students' work later in the paper.

### Epistemic Resources

Epistemic resources deal with how students perceive the nature of the knowledge under current consideration. Do they see scientific knowledge as fixed and absolute or as being relative to one's point of view? Do they view scientific knowledge as something they can construct for themselves or as something handed down from an authority figure [1]?

Epistemic resources activate and deactivate, leading a student to adopt a certain stance towards the physics knowledge at hand. Epistemic stances are highly sensitive to context [2]. A student working on a physics problem might spend time trying to build a story explaining why a certain result happened (hence treating knowledge as personally constructed) but might then turn around and search a textbook for a rule pertaining to his question (hence treating knowledge as handed down from authority). There are detailed published examples of such in-the-moment shifts in students' reasoning [3,4].

### Framing

Framing is the (often subconscious) choice the mind makes regarding "What kind of activity is going on here?" [5]. A person cannot actively consider all her knowledge in a given situation. Some subset is selected based on the mind's (often very quick) interpretation of what kind of situation is at hand.

Framing has been studied in a wide array of academic disciplines including linguistics, sociology, art, psychology, and anthropology. All of these studies implicitly agree on the existence of what has been called "Felicity's Condition" [6]. Felicity's Condition is the unspoken premise naturally adopted by an individual that incoming information, whether it be spoken, read, observed, etc, comes from a rational source, and it is thus up to the individual to attempt to contextualize and hence interpret that incoming information. Framing is the process by which the mind attempts this contextualization and interpretation.

That is not to say this contextualization always happens flawlessly. Different individuals can certainly frame the same incoming information in different ways [5]. Miscommunications arise when two individuals frame their interaction differently, each bringing a different subset of their available resources to bear on the situation. This paper presents such an example.

How a person frames a situation has many ramifications: social, conceptual, emotional, etc. This paper will focus on the epistemic ramifications of students' framing. What do they presently see as the nature of the mathematics knowledge they are using? That is, what epistemic resources have been tapped by their current framing of their problem solving?

But how can one best identify and analyze what "epistemic resources have been tapped" by physics students? We suggest: look for the (sometimes implicit) warrants they use in their mathematical arguments.

## ARGUMENTATION

Argumentation theory is a field of research about how people build justifications and communicate them effectively to each other. Since argumentation theory will provide a lens (i.e. the idea of warrants) for observing students' epistemic framing, this paper briefly turns to an overview of several threads of argumentation research.

There are several subfields that are sometimes colloquially lumped under "argumentation theory" [7]. On one end of the continuum is what is best called *formal logic*. Studies in formal logic deal with relatively clean and straightforward methods of proof-making that can easily be decontextualized from whatever given situation is at hand. Formal logic chains like "if A then B, if B then C but not D, etc." are common to such studies.

A second branch of research [7], the one that is most often actually called "argumentation theory", includes what is often called *rhetoric*. This field of research focuses most on presenting, as opposed to having, an argument. It attempts to parse the content of a given argument into some kind of structure and often carries some sort of evaluative tone with regard to that structure. A central pillar of this field, and an important basis for this paper's analysis, is the work of Stephen Toulmin [8]. He devised an often-cited system for parsing an argument into such parts as *claims*, *data*, and *warrants*. A person will make a statement, the claim, that requires proof. They will then offer one or more relevant facts, the data. The warrant is the bridge, sometimes unspoken, that explains how the given data relates to the claim at hand. For example, I might state that Jack Nicklaus is the greatest golfer alive (claim) because he won the Masters six times (data). The relevant warrant that would link this data to that claim would be that the Masters is a very prestigious tournament that is played every year on Augusta National, one of the most difficult courses on the planet.

Many naturally occurring arguments are more nebulous in structure than an argument fitting a clean Toulmin structure. Justifications that are logically unsound are often treated as acceptable in informal, real-time situations. A third branch of research, often gathered under the label *discourse analysis* [7], concerns itself primarily with the in-the-moment patterns people employ in their speech and thought as they construct and communicate arguments.

These in-the-moment argument constructions are often verbally incomplete. They often refer to a body of knowledge that the speaker (correctly or incorrectly) assumes he shares with the listener. These flow-of-conversation arguments sometimes have holes in them that are consciously or unconsciously overlooked.

This paper's analysis incorporates elements from what van Rees calls "argumentation theory" and "discourse analysis". Toulmin's idea of a warrant is central. Tracking the warrants physics students use in their mathematical arguments provides an excellent window onto their epistemic framing. We are not concerned, however, with evaluating the students' arguments. We are concerned with looking closely at a few "in-the-moment patterns" they employ in their speech and how they relate to the students' epistemic framing of their math use in physics.

## EXAMPLE

This example comes from a video of two students, a sophomore (S1) and a junior (S2), working outside of class on a homework problem from a Mathematical Methods in Physics class. Tracking their (shifting) warrants will give access to the epistemic framing dynamic that is central to their conversation.

Their homework question asks them to use $W = \int_A^B \vec{F} \cdot d\vec{r}$ to calculate the work done by the gravitational force from a mass, *M*, at the origin on a smaller mass, *m*, that moves from (1,1) to (3,3) along two dif-ferent paths. The first path goes from (1,1) to (3,3) directly while the second goes horizontally to (3,1) and then vertically to (3,3). They have decided to ignore the *GMm* constants and have written

$$\int_{\sqrt{2}}^{3\sqrt{2}} \frac{1}{r^2} dr = \int_{1}^{3} \frac{1}{y^2+9} dy + \int_{1}^{3} \frac{1}{x^2+1} dx \quad (1)$$

While the integral on the left for the radial path is correct within a minus sign, the integrals for the other path are not. There is a missing cosine term from the dot product definition of work. The students have also drawn a figure on the board that shows the two paths.

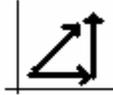

**FIGURE 1**: The students' two-path diagram.

The students discuss their expected result:

1. **S1**: What's the problem?
2.     You should get a different answer
3.     from here for this. *(Points to Fig.1)*
4. **S2**: No, no, no.
5. **S1**: They should be equal?
6. **S2**: They should be equal.
7. **S1**: Why should they be equal?
8.     This path is longer if you think about it.
    *(Points to Fig.1)*
9. **S2**: Because force, err,
10.    because work is path independent.
11. **S1**: This path is longer, so it should have,
12.    this number should be bigger than
   *(Points to Fig.1)*
13. **S2**: Work is path independent. If you
14.    go from point A to point B, doesn't matter
15.    how you get there, it should take
16.    the same amount of work.
17. **S1**: OK, that's assuming Pythagorean
18.    Theorem and everything else add[s].
19.    Well, OK, well is this—what was the
20.    answer to this right here? *(Points to Eq.1)*
21.    What was that answer?
22. **S2**: Yeah, solve each integral numerically.
23. **S1**: Yeah, what was that answer?
   *(They turn to a computer with Mathematica)*

Lines 1 to 6 introduce the crux of the episode. S1 thinks there should be different amounts of work done on the small mass along the two different paths. S2 believes the work done should be the same.

S1 offers a justification for his claim in lines 7 and 8 when he challenges S2's same-work assertion. "This path is longer if you think about it." The mathematical definition of work, $W_{A \to B} = \int_{A}^{B} \vec{F} \cdot d\vec{r}$, is essentially "force times distance". Since the two-part path from A to B is physically longer than the direct route, it seems to follow that more work is done along the longer path.

In formal argumentation theory language [8], S1 is "claiming" that more work is done along the two-part path, and he offers the "data" that the two-part path is longer. There is an unspoken "warrant" that connects his data to his claim: mathematics should align with the physical systems under study in a physics class. The goodness-of-fit between the math at hand and the physical system attests to the validity of one's conclusions. The work formula seems to say "force times distance" to S1. The two-part path has more "distance", and S1 thus draws justification for his answer.

Students' warrants offer excellent evidence for their epistemic framing of their math use. S1's warrant thus suggests he is framing his activity as *Physical Mapping*. S1 has currently activated epistemic resources that view mathematics knowledge as encoding physical ideas.

S2 not only has a different answer than S1, but he is also framing his use of mathematics in a different way. S2 claims that the work done on the small mass should be the same along the two paths "because work is path independent" (lines 9 and 10). His data is the (incompletely recalled) theorem: "work is path independent". The unspoken warrant that S2 is relying on concerns the common use of rules and definitions in math and physics: sometimes previous results are simply taken as givens for speed and convenience. S2 is framing their math use as *Invoking Authority*. He is thinking with epistemic resources that see mathematics as a system for packaging and quoting rules and previous results. Such a framing is not necessarily naïve. Even expert physicists don't necessarily start every argument from absolute first principles every time.

After hearing S2's counterargument, S1 repeats himself. In lines 11 and 12, he restates his longer-path justification and again points to the relevant features of the diagram they had previously drawn on the board. S2 responds by restating "work is path independent" in line 13 and again, in a slightly different way, in lines 14 to 16.

The most important observation in this clip is that S1 and S2 are disagreeing over much more than the answer itself. Explicitly, they are debating whether or not more work is done along the longer path. Implicitly, they are arguing over the most useful way to frame their present use of mathematics. S1 never explicitly says, "Please respond to my claim in a way that maps our math to some detail of the physical situation I may have overlooked". His phrasing and gesturing in his initial argument (lines 7 and 8) implies this framing request, though.

When S2 responds with his rule citation, he is not merely arguing for a different answer. He is pushing for a different type of warrant for judging the validity of a given answer. S2's Invoking Authority framing may have even prevented him from really hearing what S1 was saying. S1's framing request may have passed by S2 unnoticed because he was too caught up in the subset of all his math resources that his Invoking Authority framing had activated within his mind. At any rate, S2 responds in lines 9 and 10 with a different type of justification than what S1 was expecting.

When S1 repeats himself in lines 11 and 12, he is implicitly repeating his bid for a Physical Mapping framing. One can imagine a situation when S2's Invoking Authority justification would simply be accepted without incident, but here it did not align with S1's present framing. S2 does not respond to this reframing request and repeats his answer as he remains with his Invoking Authority framing.

There is thus an intense framing argument going on under the surface of this debate. Sensing that he is not making any headway in the framing battle, S1 now moves to shift both himself and S2 into a third framing.

S1 moves to reframe the discussion in lines 19 to 21. He points to the integrals they've written and asks, "Well, OK…what was the answer to this right here? What was that answer?" He is calling for someone to evaluate each of their expressions for the work so that he can compare the numeric results. This argument relies on another kind of warrant. Mathematics provides one with a standardized, self-consistent set of manipulations. Performing a computation, or having a computer do it for you, according to these rules will give a valid, trustable result. S1 is moving to reframe their math use as *Calculation*. Epistemic resources associated with this framing see mathematics as a system of algorithmic processes.

Even though S1 doesn't explicitly detail the new warrant he is proposing, S2 is quick to tune into it. He immediately responds, "Yeah, solve each integral numerically" (line 22). Compare this successful epistemic frame negotiation with the struggle earlier in this snippet. Lines 1 to 16 had S1 pushing for Physical Mapping while S2 lobbied for Invoking Authority. Both stuck to their positions, resulting in an inefficient conversation. Neither was accepting what the other was trying to say. Lines 19 to 22 have S1 and S2 agreeing, for the moment, on what type of mathematical justification should count.

## CLOSING DISCUSSION

The sample student conversation given above contains a series of epistemic framing pushes and pulls. They work to find a common ground, a shared perspective of what type of argument would be appropriate.

We do not claim that the three epistemic framings named in this episode represent rigidly compartmentalized modes of thought. We have often observed mixings, like a student using a calculation scheme to back up a physical intuition. Our main point here has been to illustrate how tracking the (sometimes implicit) warrants students use is an excellent way to gather evidence of their epistemic framing of their math use. Epistemic framing issues are often a powerful (if sometimes implicit) driving force behind students' speech and thought.

## ACKNOWLEDGMENTS

The authors wish to thank the Physics Education Research Group at the University of Maryland for their important contributions. This work was supported by NSF grants DUE 05-24987 and REC 04-40113 and a Graduate Research Fellowship.